\documentclass{nime-alternate} 
\usepackage{diagbox}
\usepackage{anonymize} 		   

\usepackage[utf8]{inputenc}
\usepackage{svg}
\usepackage{amsmath}
\usepackage{stfloats} 

\usepackage{etoolbox}
\newtoggle{svg}
\newtoggle{png}

\toggletrue{png}

\begin{document}


\title{Enhancing DMI Interactions by Integrating Haptic Feedback for Intricate Vibrato Technique}

\label{key}
%

\numberofauthors{4}

\author{
\alignauthor
\anonymize{Ziyue Piao}\\
       \affaddr{\anonymize{IDMIL, CIRMMT, McGill University}}\\
       \affaddr{\anonymize{Montreal, Quebec, Canada}}\\
       \email{\anonymize{ziyue.piao@mail.mcgill.ca}}
\alignauthor
\anonymize{Christian Frisson}\\
       \affaddr{\anonymize{IDMIL, McGill University}}\\
       \affaddr{\anonymize{Montreal, Quebec, Canada}}\\
       \email{\anonymize{christian.frisson@gmail.com}}
\alignauthor
\anonymize{Bavo Van Kerrebroeck}\\
       \affaddr{\anonymize{IDMIL, CIRMMT, McGill University}}\\
       \affaddr{\anonymize{Montreal, Quebec, Canada}}\\
       \email{\anonymize{bavo.vankerrebroeck@mail.mcgill.ca}}
\and  
\alignauthor
\anonymize{Marcelo M.Wanderley}\\
       \affaddr{\anonymize{IDMIL, CIRMMT, McGill University}}\\
       \affaddr{\anonymize{Montreal, Quebec, Canada}}\\
       \email{\anonymize{marcelo.wanderley@mcgill.ca}}
}


\maketitle

\begin{abstract}
This paper investigates the integration of force feedback in Digital Musical Instruments (DMI), specifically evaluating the reproduction of intricate vibrato techniques using haptic feedback controllers. We introduce our system for vibrato modulation using force feedback, composed of Bend-aid (a web-based sequencer platform using pre-designed haptic feedback models) and TorqueTuner (an open-source 1 Degree-of-Freedom (DoF) rotary haptic device for generating programmable haptic effects). We designed a formal user study to assess the impact of each haptic mode on user experience in a vibrato mimicry task. Twenty musically trained participants rated their user experience for the three haptic modes (Smooth, Detent, and Spring) using four Likert-scale scores (\textit{comfort}, \textit{flexibility}, \textit{ease of control}, and \textit{helpfulness} for the task). Finally, we asked participants to share their reflections. Our research indicates that while the Spring mode can help with \textit{Light vibrato}, preferences for haptic modes vary based on musical training background. This emphasizes the need for adaptable task interfaces and flexible haptic feedback in DMI design.

\end{abstract} 
\keywords{Rotary force feedback, Haptics, Vibrato, TorqueTuner}

\ccsdesc[500]{Applied computing~Sound and music computing}
\ccsdesc[500]{Hardware~Haptic devices}

\printccsdesc

\section{Introduction}

\begin{figure}[h]
  \centering
    \iftoggle{svg}{%
        \includesvg[
            pretex=
                \fontsize{6}{5}\selectfont 
                \sffamily
                ,
            width=1.10\linewidth
        ]{Figure1_v2.svg}
    }{}
    \iftoggle{png}{%
        \includegraphics[width=\linewidth]{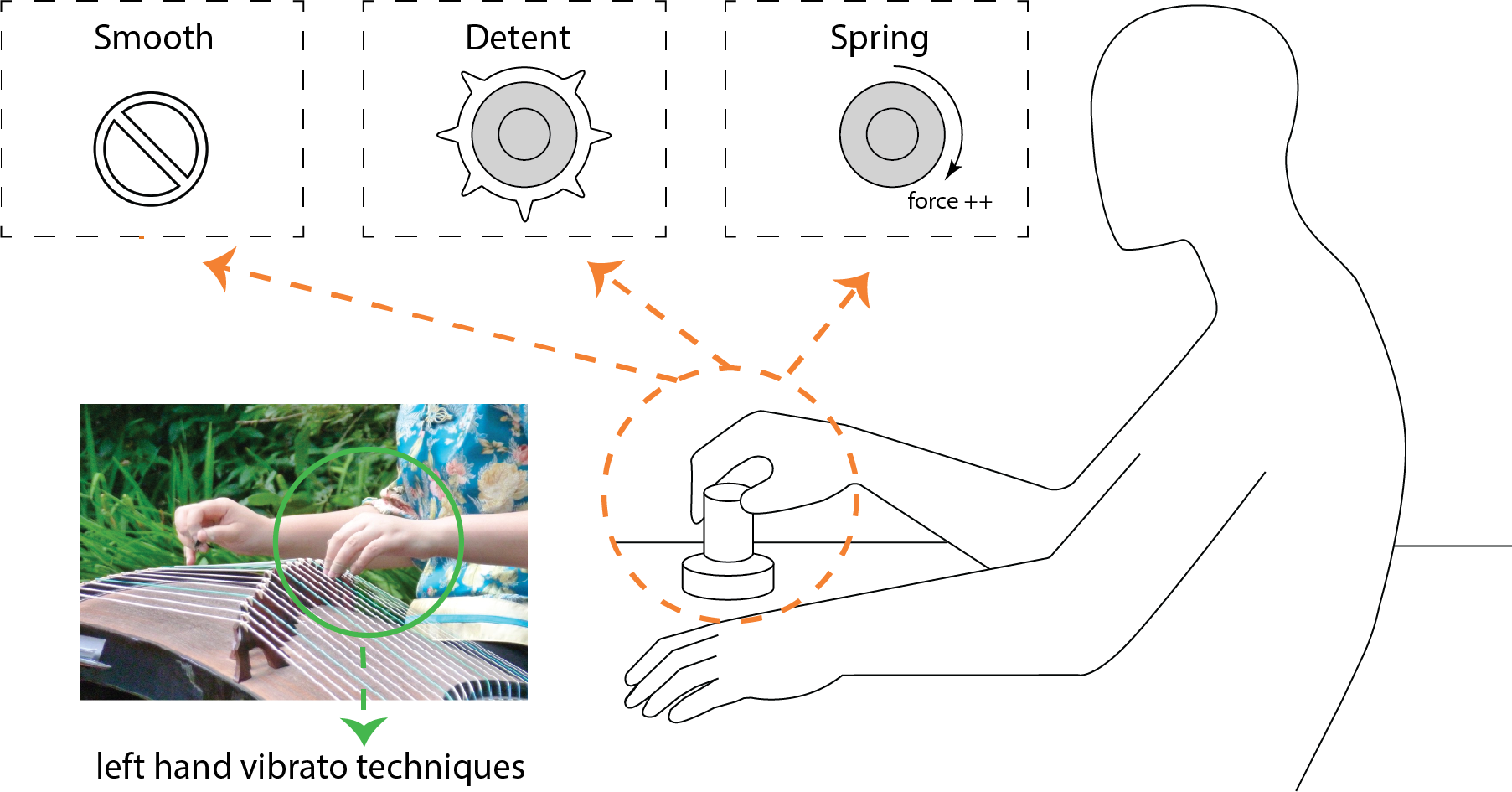}
    }{}
  \caption{A picture of a person interacting with a Guzheng, embedded in a larger sketch of a person interacting with a haptic DMI inspired by the Guzheng with three haptic modes}
  \label{fig:figure1}
\end{figure}

Adding force feedback to DMIs could potentially enable more complex and expressive sounds to knobs in traditional musical devices, addressing the challenges and opportunities identified by \cite{Frisson2023FFM} for force-feedback digital musical instruments. In this paper, we explore the application of three force feedback modes to enhance the expression of Guzheng vibrato technique through a programmable knob, because of the ubiquity of rotary controls in musical hardware. We utilized the embedded device TorqueTuner to provide dynamic and modifiable rotary force feedback to enable precise control over complex vibrato techniques \cite{Kirk2020NIME}. We also developed Bend-aid, a web-based MIDI editing tool that integrated with the TorqueTuner, enabling musicians to apply specific haptic feedback to pitch-related techniques. To assess the effectiveness of our haptic feedback system, we conducted a user study focusing on the replication of the Guzheng's \textit{Light vibrato} technique. Participants were exposed to the three force feedback settings. We evaluated their experiences through a subjective Likert-scale rating and gathered reflections on their perceived effectiveness of the force feedback.

\section{Related work}

Musical instrument players often rely on physical feedback to refine their techniques and achieve desired sound and expressiveness \cite{Young2018MH}. Touch-free instruments like the Theremin, when equipped with haptic feedback, have been shown to enhance playability \cite{o2001playing}. Researchers are actively exploring the integration of haptic feedback in digital interfaces to improve performance control, expressivity, and user experience \cite{papetti2018musical}.

The majority of efforts in adding haptic feedback to virtual instruments aim to replicate the tactile sensations of acoustic musical instruments. Examples include virtual drums with vibrotactile actuators \cite{nichols2002vbow} and virtual pianos with mid-air haptic displays \cite{hwang2017airpiano}. For string instruments, haptic feedback is achieved through robotic arms and vibrotactile actuators to simulate the actions of plucking, bowing, and rubbing \cite{onofrei2023perceptual, andrea2019no}. 

In music production, research focuses on enabling control over musical features through force feedback. Tools like the Haptic Wave, designed for visually impaired producers and engineers \cite{Tanaka2016CHI}, demonstrate the effectiveness of haptic feedback in audio mixing \cite{Merchel2010AES} and audio effects editing \cite{De2021DAFx}. Projects such as SoundFORMS \cite{Colter2016CHI} and HaptEQ \cite{Karp2017AM} allow users to interact with and alter synthesized waveforms using gestures, highlighting force feedback's potential to refine musical control in production.

Rotary force-feedback controllers are commonly used in vehicular instrument control, robot control, and media editors \cite{De2021DAFx}. Rotary control, widely used in traditional musical instruments like pitch bend wheels and knobs, often lacks force feedback or provides only constant torque. Tools like THE PLANK address this gap by providing one-axis force feedback for precise and rapid controls in live performances \cite{Verplank2002NIME}. Similarly, the D’groove intelligent DJ system utilizes a haptic turntable for controlling digital audio playback with added haptic feedback \cite{Beamish2003ICAD}. TorqueTuner \cite{Kirk2020NIME,TorqueTunerHAID2022}, a project using a DC motor for programmable force feedback in digital musical instruments, shows promise but lacks experimental validation of its effectiveness.

\section{Interface Design}

\begin{figure}[h]
  \centering
  \includegraphics[width=\linewidth]{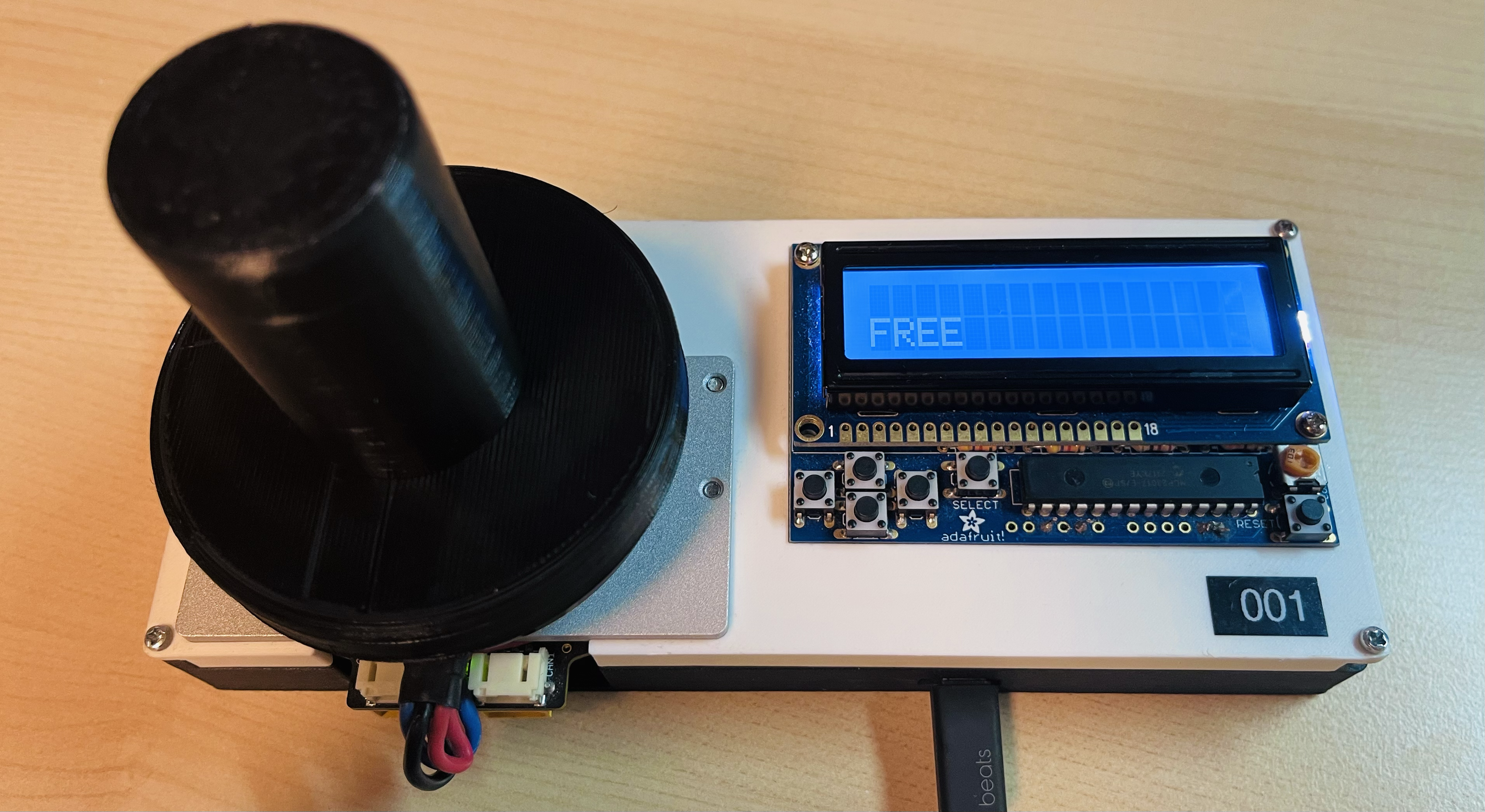}
  \caption{TorqueTuner with our easy-to-hold grip}
  \label{fig:torquetuner}
\end{figure}


We created Bend-aid, a visual interface, to explore our key question: Does force feedback improve the mimicry of intricate Guzheng vibrato? Bend-aid enables bi-directional communication with TorqueTuner, facilitating the editing of pitch contours using varied force feedbacks. We designed and 3D-printed a grip for the TorqueTuner, facilitating rapid shifting gestures essential for executing \textit{Light vibrato}, as illustrated in Figure~\ref{fig:torquetuner}. In this section, we describe various aspects of the Bend-aid's interface design, including visual interface design and haptic modes design. We designed two haptic modes with force feedback (\textit{Spring} mode and \textit{Detent} mode) and one without force feedback as the baseline (\textit{Smooth} mode).

\subsection{Visual Interface Design}

\begin{figure*}[t]
  \centering
    \iftoggle{svg}{%
        \includesvg[
            pretex=
                \sffamily
                ,
            width=\linewidth
            ]{Bend-aid-ui.svg}
    }{}
    \iftoggle{png}{%
        \includegraphics[width=\linewidth * 3 / 4]{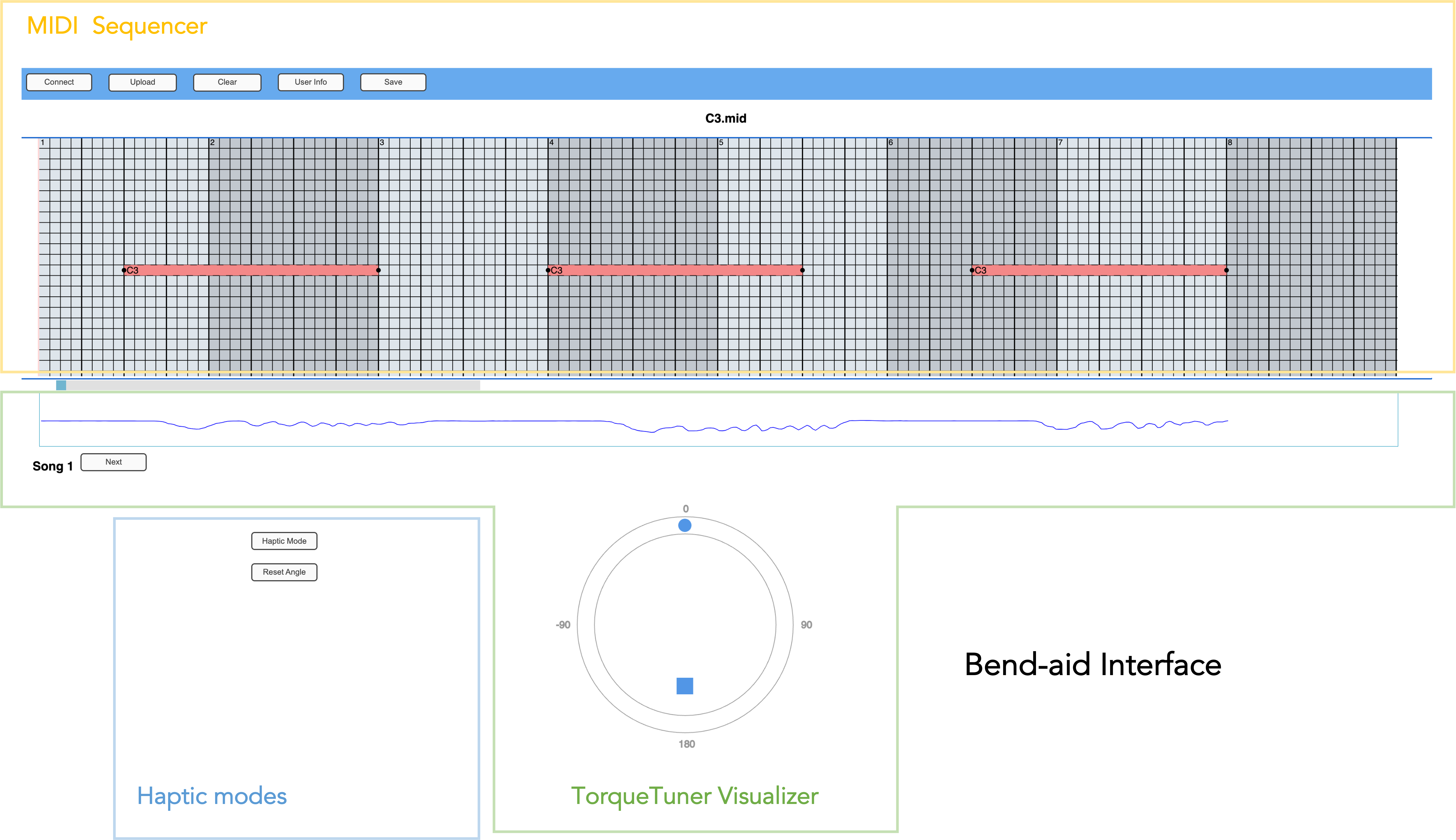}
    }{}
  \caption{UI design of Bend-aid system.}
  \label{fig:bend-aid-ui}
\end{figure*}

Bend-aid is a web-based interface, as shown in Figure~\ref{fig:bend-aid-ui}, designed for editing MIDI notes and incorporating various rotary force feedbacks to facilitate pitch bending. Gesture data can be inputted through TorqueTuner, allowing interaction with MIDI notes using a mouse and keyboard. The real-time angle of TorqueTuner is mapped to pitch value in this study. 

\subsubsection{MIDI sequencer}
We integrated a MIDI sequencer based on an embeddable piano roll interface called powerPianoRoll \cite{powerPianoRoll} to replicate the typical MIDI editing interaction in Digital Audio Workstations (DAW). MIDI sequencer facilitates the design of our user study and potentially makes the editing process intuitive. The audio synthesizer is implemented using the Tone.js framework and the Guzheng sampler \cite{GuzhengSample}. The sequencer supports several MIDI editing functions similar to commonly used DAWs like Logic Pro and Ableton Live, including:

\begin{enumerate}
    \item Double click on an empty cell to add a note.
    \item Click to highlight a selection. 
    \item Drag a note to move it.
    \item Drag the beginnings and ends of a note to resize it.
\end{enumerate}

The MIDI sequencer supports editing long pieces by dragging the slider beneath it and can handle notes from low C1 up to high G8. Notes can be entered manually, the upload'' button allows for importing a MIDI file from a local folder, and the clear'' button can be used to remove all notes on the sequencer canvas.

\subsubsection{Data communication}

\begin{figure*}[t]
  \centering

        \includegraphics[width=\linewidth * 5 / 7]{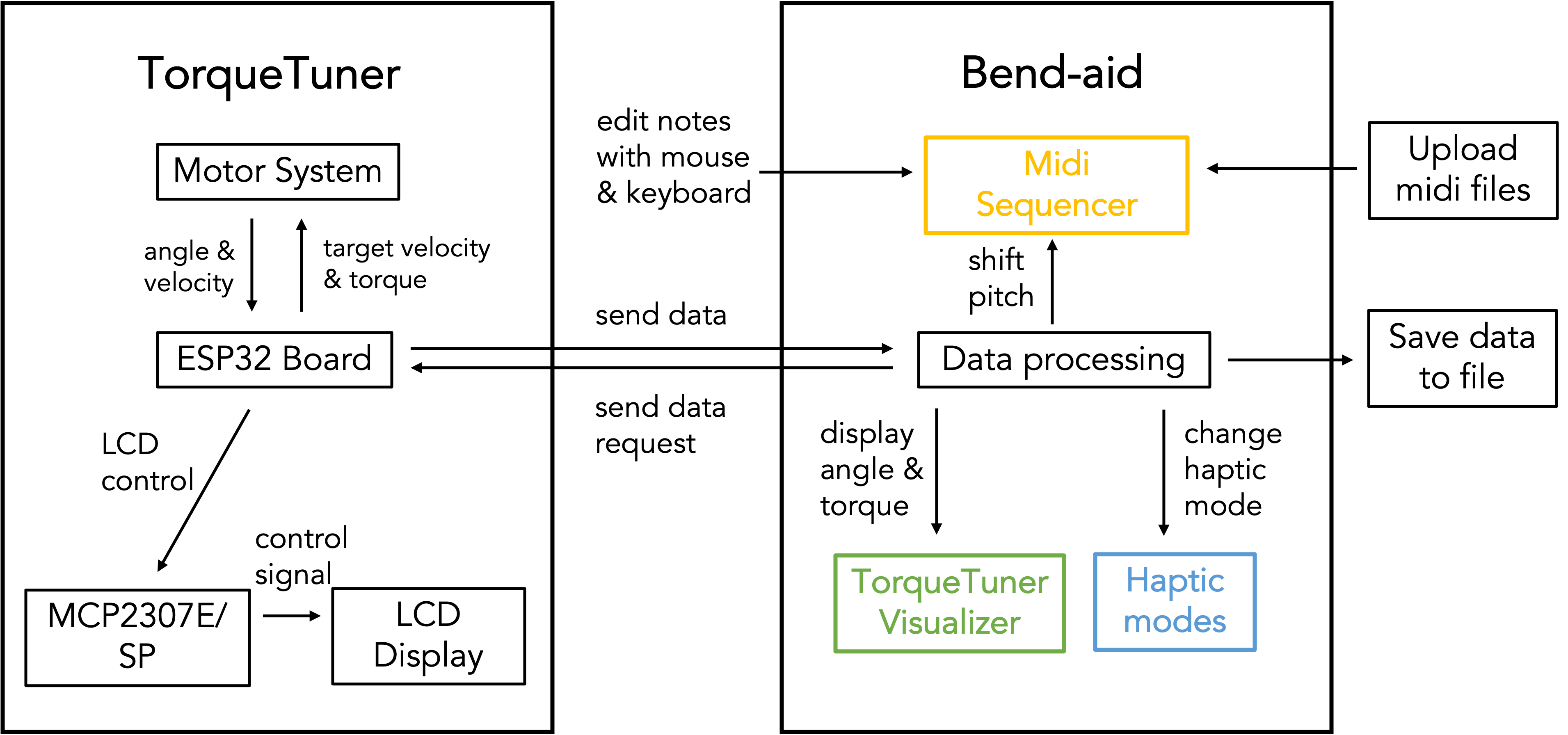}
  \caption{Block diagram of TorqueTuner and Bend-aid system}
    \label{fig:diagram}
\end{figure*}
The overall communication diagram of TorqueTuner and Bend-aid is shown in Figure~\ref{fig:diagram}. Once TorqueTuner is connected to the computer via a USB serial port, we can press the ``connect'' button and select the serial port of the TorqueTuner. The TorqueTuner can then communicate with Bend-aid. In each loop, the TorqueTuner prints real-time torque, angle, velocity, and mode name to the serial port, which Bend-aid reads to update its variables. The visual interface displays the haptic mode on the interface, the LCD of the TorqueTuner, and the real-time torque and angle of the knob. When changing haptic modes on the Bend-aid, a stream of the new mode is sent to the TorqueTuner to change its mode and display the new haptic mode on its LCD screen.

\subsubsection{TorqueTuner visualizer}
A knob visualizer was designed in Bend-aid to enable observation of the knob's status on TorqueTuner. The visualizer smoothly displays the angle with a small blue circle rotating around a larger circle, which serves as a pointer and shows a bar-like meter indicating torque value. When changing haptic modes, the current knob status serves as the zero point, and the angle and torque are reset until there is another interaction with TorqueTuner. Furthermore, we have plotted the angle contour below the MIDI Sequencer, which can be saved as a file for data analysis. 

\subsection{Haptic Modes Design}
The haptic modes in Bend-aid have been designed based on the commonly used rotary input devices in music hardware. Next to the three haptic modes (Smooth, Detent, and Spring), we designed and tested some haptic modes that were excluded after an initial pilot study. The Free mode maps a constant value to the Torque scale. The Vibrato mode, when the knob is turned or stopped at one angle, it continuously rotates left and right with a small motion, and the torque changes in a sinusoidal period. We also explored the Magnet, Friction, Inertia, and Spin haptic modes by adjusting their 
parameters from the previous version of TorqueTuner \cite{Kirk2020NIME}. Finally, we gave up using these haptic modes during iterations of the initial pilot study because of limited user experience compared with the other three modes. 

Because the \textit{Light vibrato} require subtle and fast changes between pitch values, we found that the participant could not finish the mimicry task when the target pitch range was mapped into a large knob angle degree. We decided to focus instead on the force feedback design between 0-90 degrees 
with three haptic modes (Smooth, Detent, and Spring).

\subsubsection{Smooth}
The Smooth mode has negligible resistance when twisting and serves as the control condition in the following experiment to compare outcomes with and without haptic force feedback.

\subsubsection{Detent}
The Detent mode provides a mechanical click when the knob is turned to a certain position. This mode is commonly found in musical knobs with discrete settings, like volume control. In each 45 degree segment, the torque exponentially increases, opposing the turning direction. A transient change in torque direction at the position of the Detent effectively results in a haptic click. To facilitate the mimicry task, we increased resistance at the opposing torque point and make the point directly linked to the target highest pitch in \textit{Light vibrato}. We aim at creating a boundary hint, enabling quicker angle adjustments for rapid pitch changes.

\subsubsection{Spring}
The Spring mode provides resistance that increases as the knob is turned to simulate the haptic feeling of pressing a string. This can be applied to knobs adjusting continuous parameters, such as audio effect settings and the pitch bend wheel. We tested exponential and linear relationships between torque and angular velocity and finally chose the linear relationship for a more intuitive interaction similar to pressing a string based on Hooke's Law. 

\section{User Study}
This study evaluated the effects of three force feedback modes (Smooth, Detent, and Spring) on mimicking the \textit{Light vibrato} technique in Guzheng. First, participants completed a pre-experiment questionnaire on their musical background. Using a within-subjects design, participants then attempted to memorize and replicate \textit{Light vibrato} with Bend-aid under three force feedback conditions. The sequence of feedback conditions was counterbalanced to avoid order effects. After attempts in each haptic mode, participants rated their experience, focusing on \textit{comfort}, \textit{ease of control}, \textit{flexibility}, and \textit{helpfulness}. Additionally, qualitative feedback was collected through a post-study interview to delve deeper into their reflections of the task.

\subsection{Participants}
Preliminary tests with both musically trained and untrained individuals indicated that musical background significantly influenced task performance. Hence, for the main study, we recruited 20 musically trained participants (7 females, 13 males), compensating them \$10 for their 30-minute participation.

\subsection{Procedure}
The experimenter first introduced the study procedure to the participants and collected the consent form. Before the mimicry task, the participant was required to fill out a questionnaire on their musical background (Table \ref{tab:questionnaire}). Then, the experimenter connected the TorqueTuner to a testing laptop (MacBook Pro 13), opened Bend-aid interface, and helped the participant put on headphones (AKG K240 Studio). After introducing the Bend-aid interface, the experimenter gave the participant the opportunity to become familiar with the pitch control using TorqueTuner. The experiment session followed. In each activity, the participant listened to a short piece of audio with three repetitive C3 notes with \textit{Light vibrato} as Figure \ref{fig:bend-aid-ui} shows (each lasting six quarter notes in the 4/4 time signature at 120 bpm). The experimenter then asked the participant to mimic the sound in Bend-aid. Participants were allowed to try multiple times until they were satisfied with their performance. Once the participant finished the vibrato mimicry in one haptic mode, they completed a Likert-scale questionnaire asking for their experienced comfort, and the device’s flexibility, ease of control, and helpfulness to the task (Table \ref{tab:likert}). Subsequently, the participants finished another two sessions with new haptic modes and rated their experience using the Likert Scale. Finally, the experimenter conducted a semi-structured post-experiment interview (Table \ref{tab:questionnaire}).

\begin{table}[h!]
\centering
\begin{tabular}{|p{0.25\linewidth}|p{0.65\linewidth}|}
\hline
\textbf{Pre-experiment Questionnaire} & \textbf{Questions} \\
\hline
Q1 & Do you have experience in learning musical instruments (including singing)? \\
Q2 & How many instruments can you play? Please list two of your most proficient instruments and how many years you played them. \\

\hline
\textbf{Post-experiment Interview} & \textbf{Questions} \\
\hline
Q1 & Which haptic modes do you like the most and the least? Why? \\
Q2 & Do you have any problem using the interface to control vibrato? \\
\hline
\end{tabular}
\caption{Pre- and post-experiment questions}
\label{tab:questionnaire}
\end{table}

\begin{table}[h!]
\centering
\begin{tabular}{|p{0.25\linewidth}|p{0.65\linewidth}|}
\hline
\textbf{Likert-scale ratings} & \textbf{Explanations} \\
\hline
Comfort & Evaluates how physically pleasant and stress-free the haptic mode feels to the user during the task.\\
Ease of control & Measures the user's flexibility to adjust the angle in the haptic mode to suit the vibrato technique. \\
Flexibility & Assesses how easily users can use TorqueTuner to accomplish tasks in the haptic mode. \\
Helpfulness & Measures how much the haptic mode helps the user improve their performance and experience in the mimicry task. \\
\hline
\end{tabular}
\caption{Explanation for the four Likert-scale ratings}
\label{tab:likert}
\end{table}

\section{Results}
This section presents the findings from our user study, including subjective ratings and qualitative responses.

\subsection{Subjective Ratings}

\begin{figure*}[t]
  \centering
    \iftoggle{svg}{%
        \includesvg[
            pretex=
                \sffamily
                ,
            width=1.175\linewidth 
            ]{image-source/subjective-rating-violin.svg}
    }{}
    \iftoggle{png}{%
        \includegraphics[width=\linewidth]{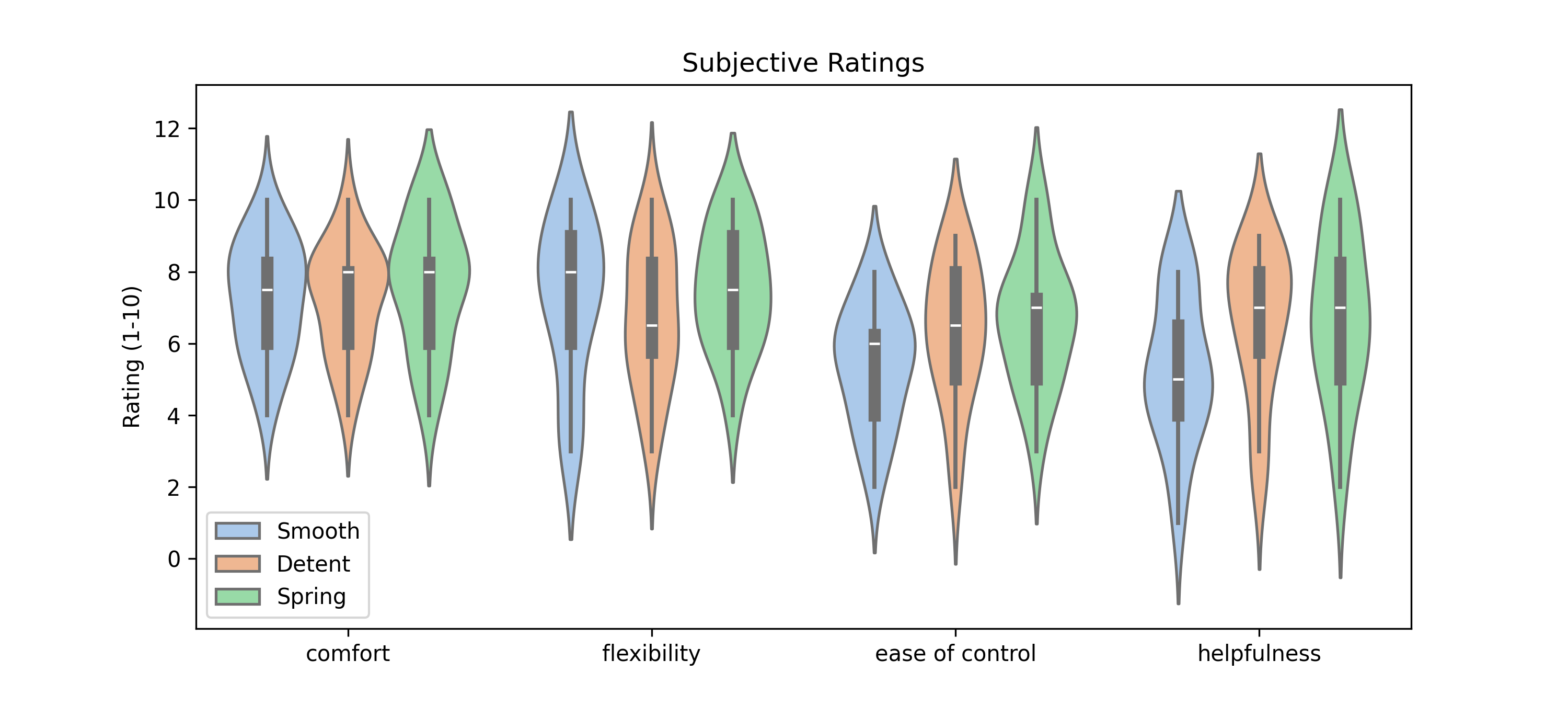}
    }{}
  \caption{Likert-scale ratings from 20 participants, including means and standard deviations for \textit{comfort}, \textit{flexibility}, \textit{ease of control}, and task \textit{helpfulness}}
  \label{fig:subjective-rating}
\end{figure*}

We hypothesize that the three haptic modes have varying impacts on participants' subjective ratings. To test this, we collected data from 20 participants across four categories (\textit{comfort}, \textit{ease of control}, \textit{flexibility}, and \textit{helpfulness}) on a 10-point Likert scale. We also calculated the average score for each haptic mode across the four categories, identifying the mode with the highest average as the user's preferred mode. In Figure \ref{fig:subjective-rating}, we visualized the Likert-scale ratings, including means and standard deviations for the four categories. We utilized the violin plot to better visualize statistical distributions than with only box plots \cite{matejka2023violin}.

We ran one-way repeated measures ANOVA on the four categories and the following analysis is based on the new statistics \cite{cumming2013understanding}. 

For \textit{comfort}, the ANOVA revealed no significant differences among three haptic modes ($F(2, 57) = 0.347673, p = 0.707817, \eta^{2} = 0.012$), with a negligible effect size suggesting minimal variation attributable to three haptic modes. The overlapping confidence intervals (CI) among Detent (CI: 6.48 to 7.92), Smooth (CI: 6.49 to 8.01), and Spring (CI: 6.76 to 8.44) reinforce the conclusion of no substantive difference in perceived \textit{comfort}. \textit{Flexibility} ratings similarly showed no significant effect of haptic mode condition ($F(2, 57) = 0.725, p = 0.489, \eta^{2} = 0.025$). Confidence intervals for Detent (CI: 5.83 to 7.67), Smooth (CI: 6.15 to 8.25), and Spring (CI: 6.70 to 8.30) further confirm the lack of significant differences across haptic conditions for \textit{Flexibility}. These findings suggest that incorporating force feedback into knobs does not negatively affect \textit{comfort} or \textit{flexibility}, even though turning the knob with haptic feedback requires more effort.

\textit{Ease of control} ($F(2, 57) = 2.556, p = 0.086, \eta^{2} = 0.082$) and \textit{helpfulness} ($F(2, 57) = 2.845, p = 0.066, \eta^{2} = 0.091$) nearly reached significance, suggesting effects of haptic conditions. For \textit{ease of control}, the CIs reveal a noticeable gap between Smooth (CI: 4.62 to 6.18), particularly between Smooth and Spring (CI: 5.79 to 7.51). For \textit{helpfulness}, the CIs for Smooth (CI: 4.29 to 6.21) shows a similar gap with that of Spring (CI: 5.63 to 7.77). The results suggest that for participants, the Spring mode is better than Smooth mode in making the knob ease of control and helping with finishing the task.

We used the chi-square statistic complemented by Cramér's V \cite{cramer1948stat} to examine whether musical background influences participants' preferences for haptic feedback. Participants were broadly categorized based on their musical experience into wind, and string instrument players, omitting other categories such as piano, voice, and percussion due to either their prevalence or scarcity. Our findings reveal a statistically significant association between wind and string instrument players' preferences ($\chi^2(2) = 12.898, p < 0.05$) with Cramér's V indicating a strong relationship ($V = 0.568 > 0.5$). This suggests that the type of musical instrument experience significantly influences haptic feedback preferences, particularly distinguishing wind from string players. Specifically, 77.8\% of wind instrument players preferred the \textit{Spring} mode, while 62.5\% of string players preferred the \textit{Detent} mode.

\subsection{Qualitative Responses}
After the quantitative experiment, we gathered reflections from all participants via a post-experiment interview.

Two participants, one with experience in electric guitar and another in percussion, emphasized their reliance on tactile feedback when interacting with Bend-aid and TorqueTuner. The guitar player noted that \textit{``the Spring mode is the most similar mode to the tactile sensation of strings, so I can easily apply the mode in the vibrato task''}. Two piano performers shared a consistent experience, noting: \textit{``Initially, the haptic sensation of the Spring mode was unfamiliar for me. However, once accustomed to it, it became the most intuitive mode''}. Additionally, two participants favored the Spring mode because of its assistance with the variation towards the end of the vibration. Two participants expressed a preference for interacting with haptic feedback at lower force levels, leading them to choose the Smooth mode as their favorite. However, one electric guitar player found the Smooth mode challenging for executing rapid vibrato movements due to the lack of haptic feedback. Given that many instruments use linear motor gestures for pitch control, a participant with trombone experience proposed that \textit{``applying the results into linear force feedback may make the interaction easier and more similar with playing the Guzheng"}.

\section{Discussion}

We explore the outcomes of incorporating diverse force feedback mechanisms into Guzheng vibratos, as well as the implications for integrating tactile sensations into future DMIs.

\subsection{Haptic Feedback and User Experience}
The subjective ratings underscore incorporating force feedback into knobs does not adversely affect \textit{comfort} or \textit{flexibility} in the three haptic modes. As for \textit{ease of control} and \textit{helpfulness}, the Smooth mode and Spring mode has a significant difference. The result shows that Spring mode can improve the perceived performance in the mimicry task, suggesting that involving physical feedback may enhance intricate effectiveness of DMIs without compromising user experience in flexibility and comfort.

Wind instrument players preferred the Spring mode, while string players were less enthusiastic. This difference may be due to string players' increased sensitivity to the slight mismatches between the force feedback from the TorqueTuner and the feel of real strings. Musicians' varied experiences with acoustic instruments shape their expectations and preferences for DMIs. This variation suggests that a one-size-fits-all approach to haptic design in DMIs may be less effective than a more tailored strategy that considers the specific tactile nuances associated with different instrumental backgrounds.

\subsection{Limitations and Future DMI Interaction}
The study can be further explored in these aspects: 1) the need for quantitative analysis by comparing pitch contours and angle time-series data with Bend-aid's ``save'' function; 2) comparing the effectiveness of linear force feedback and rotary force feedback in the mimicry task; 3) enhancing user experience by adjusting force feedback based on participants' grip strength; 4) investigating more haptic modes customized to musical backgrounds to identify similarities and differences compared to our findings.

The preferences based on musical background further emphasize the need for customizable haptic feedback within digital musical interfaces. The demonstrated preference for particular haptic modes and their impact on performance underscores the value of adjustable, background-specific force feedback in improving both user experience and musical expressiveness. Furthermore, incorporating force feedback insights from traditional instruments offers a promising approach to immersive and natural interactions with DMIs.

\section{Conclusion}
This study delved into enhancing DMIs with force feedback for replicating Guzheng's vibrato techniques, utilizing our developed Bend-aid and TorqueTuner. By examining three haptic modes (Smooth, Detent, and Spring), we gained insights into their impact on user experience, particularly in terms of \textit{comfort}, \textit{ease of control}, \textit{flexibility}, and \textit{helpfulness} to the vibrato mimicry task. We discovered a marked \textit{helpfulness} and \textit{ease of control} for Spring modes over the Smooth mode. Additionally, musicians' backgrounds influenced haptic preferences, revealing differences between wind and string instrument players. We emphasize the necessity for DMIs to feature customizable haptic feedback that aligns with the specific requirements of the music and the musician. Our findings suggest promising avenues for DMI innovation, such as incorporating force feedback insights from traditional instruments and tailoring feedback intensity to individual users' physical capabilities. These developments potentially further enhance the intuitiveness of force-feedback digital musical instruments.

\section{Ethical Standards}
All data including questionnaire ratings and post-experiment reflections during the experiment was approved by the \anonymize {McGill University Research Ethics Board Office (file number 23-06-074)} and consent by participants.

\section{Acknowledgement}
The authors extend their gratitude to \anonymize{João Tragtenberg, Albert-Ngabo Niyonsenga, Fausto Borem, Maxwell Gentili-Morin, and Lejun Min} for their enthusiastic and invaluable assistance and suggestions. Additionally, we thank all the participants for their valuable contributions and feedback. This work was partially supported by the \anonymize{NSERC Discovery Grant \#206719}.
%

\bibliographystyle{ieeetr}
\bibliography{nime-references}

\begin{thebibliography}{10}

\bibitem{Frisson2023FFM}
C.~Frisson and M.~M. Wanderley, ``Challenges and opportunities of force feedback in music,'' {\em Arts}, vol.~12, no.~4, 2023.

\bibitem{Kirk2020NIME}
M.~Kirkegaard, M.~Bredholt, C.~Frisson, and M.~Wanderley, ``Torquetuner: A self contained module for designing rotary haptic force feedback for digital musical instruments,'' in {\em Proceedings of the international conference on new interfaces for musical expression}, pp.~273--278, 2020.

\bibitem{Young2018MH}
G.~W. Young, D.~Murphy, and J.~Weeter, ``A functional analysis of haptic feedback in digital musical instrument interactions,'' {\em Musical Haptics}, pp.~95--122, 2018.

\bibitem{o2001playing}
M.~S. O'Modhrain, {\em Playing by feel: incorporating haptic feedback into computer-based musical instruments}.
\newblock Stanford University, 2001.

\bibitem{papetti2018musical}
S.~Papetti and C.~Saitis, {\em Musical haptics}.
\newblock Springer Nature, 2018.

\bibitem{nichols2002vbow}
C.~Nichols, ``The vbow: development of a virtual violin bow haptic human-computer interface,'' in {\em Proceedings of the 2002 conference on new interfaces for musical expression}, pp.~1--4, 2002.

\bibitem{hwang2017airpiano}
I.~Hwang, H.~Son, and J.~R. Kim, ``Airpiano: Enhancing music playing experience in virtual reality with mid-air haptic feedback,'' in {\em 2017 IEEE world haptics conference (WHC)}, pp.~213--218, IEEE, 2017.

\bibitem{onofrei2023perceptual}
M.~G. Onofrei, F.~Fontana, and S.~Serafin, ``Perceptual relevance of haptic feedback during virtual plucking, bowing and rubbing of physically-based musical resonators,'' in {\em Arts}, vol.~12, p.~144, MDPI, 2023.

\bibitem{andrea2019no}
P.~Andrea, P.~Razvan, N.~C. Nilsson, N.~S. Andersson, F.~Fontana, N.~Rolf, S.~Stefania, {\em et~al.}, ``No strings attached: Force and vibrotactile feedback in a guitar simulation,'' in {\em Proceedings of the 16th Sound \& Music Computing Conference}, pp.~210--216, 2019.

\bibitem{Tanaka2016CHI}
A.~Tanaka and A.~Parkinson, ``Haptic wave: A cross-modal interface for visually impaired audio producers,'' in {\em Proceedings of the 2016 CHI Conference on Human Factors in Computing Systems}, pp.~2150--2161, 2016.

\bibitem{Merchel2010AES}
S.~Merchel, E.~Altinsoy, and M.~Stamm, ``Tactile music instrument recognition for audio mixers,'' in {\em Audio Engineering Society Convention 128}, Audio Engineering Society, 2010.

\bibitem{De2021DAFx}
Y.~De~Pra, F.~Fontana, and S.~Papetti, ``Interacting with digital audio effects through a haptic knob with programmable resistance,'' in {\em 2021 24th International Conference on Digital Audio Effects (DAFx)}, pp.~113--120, IEEE, 2021.

\bibitem{Colter2016CHI}
A.~Colter, P.~Davivongsa, D.~D. Haddad, H.~Moore, B.~Tice, and H.~Ishii, ``Soundforms: Manipulating sound through touch,'' in {\em Proceedings of the 2016 CHI Conference Extended Abstracts on Human Factors in Computing Systems}, pp.~2425--2430, 2016.

\bibitem{Karp2017AM}
A.~Karp and B.~Pardo, ``Hapteq: A collaborative tool for visually impaired audio producers,'' in {\em Proceedings of the 12th International Audio Mostly Conference on Augmented and Participatory Sound and Music Experiences}, pp.~1--4, 2017.

\bibitem{Verplank2002NIME}
B.~Verplank, M.~Gurevich, and M.~V. Mathews, ``The plank: Designing a simple haptic controller.,'' in {\em Proceedings of the 2002 Conference on New Interfaces for Musical Expression}, pp.~33--36, 2002.

\bibitem{Beamish2003ICAD}
T.~Beamish, K.~van~de Doel, K.~MacLean, and S.~Fels, ``D'groove: A haptic turntable for digital audio control,'' in {\em Proceedings of the 2003 International Conference on Auditory Display}, 2003.

\bibitem{TorqueTunerHAID2022}
A.-N. Niyonsenga, C.~Frisson, and M.~M. Wanderley, ``{{TorqueTuner}}: A {{Case Study}} for {{Sustainable Haptic Development}},'' in {\em 11th {{Intl}}. {{Workshop}} on {{Haptic}} and {{Audio Interaction Design}}}, {{HAID}}'22, 2022.

\bibitem{powerPianoRoll}
A.~Sarwate and J.~Armitage, ``{powerPianoRoll}.'' \url{https://github.com/AvneeshSarwate/powerPianoRoll}.

\bibitem{GuzhengSample}
Cashhammn, ``{Guzheng-sample}.'' \url{https://github.com/Cashhammn/Guzheng-sample}.

\bibitem{matejka2023violin}
J.~Matejka and G.~Fitzmaurice, ``Same stats, different graphs: Generating datasets with varied appearance and identical statistics through simulated annealing,'' in {\em Proceedings of the 2017 CHI Conference on Human Factors in Computing Systems}, CHI '17, (New York, NY, USA), p.~1290–1294, Association for Computing Machinery, 2017.

\bibitem{cumming2013understanding}
G.~Cumming, {\em Understanding the new statistics: Effect sizes, confidence intervals, and meta-analysis}.
\newblock Routledge, 2013.

\bibitem{cramer1948stat}
H.~Cramér, {\em Mathematical Methods of Statistics}, ch.~Chapter 21. The two-dimensional case, p.~282.
\newblock Princeton University Press, 1946.

\end{thebibliography}

\end{document}